\documentclass[reprint,amsmath,amssymb,aps,nofootinbib]{revtex4-1}
\usepackage{dcolumn}
\usepackage{bm}
\usepackage{amsmath,amsthm,amsfonts,float,latexsym}
\usepackage[pdftex]{hyperref}
\begin{document}
\title{Imbedding a Reissner-Nordstr\"om charged mass into cosmology}
\author{Camilo Posada}
 \affiliation{%
Department of Physics and Astronomy, University of South Carolina, Columbia, SC 29208 USA.}
%
%
\date{\today}%
\begin{abstract}
We present the extension of the method of imbedding a mass into cosmology proposed by Gautreau, for the case of a Reissner-Nordstr\"om charged mass. We work in curvature coordinates $(R,T)$ where the coordinate time $T$ is measured by clocks at fixed points $R = $ const., and geodesic coordinates $(R,\tau)$ where $\tau$ is the time recorded by clocks moving along the radial geodesics. The Einstein equations are solved for a energy-momentum tensor which is composed of a cosmological fluid and an electrostatic field outside a radius $R_{b}$. Inside $R_{b}$ we have a part of cosmological fluid plus a Reissner-Nordstr\"om charged mass. We offer metrics for a Reissner-Nordstr\"om charged mass imbedded into different cosmological scenarios with zero curvature. An important consequence of our results, is that orbits will spiral when a charged mass is imbedded into a cosmological background. We found the generalized equation for the change of orbital radius, by using geodesic coordinates. Some criticism to the Newtonian calculation done by Gautreau for the orbital spiralling is discussed.  
\end{abstract}
\maketitle

\section{Introduction}
\vskip-0.05in
It's well known that the Reissner-Nordstr\"om metric describes the gravitational field due to a charged mass into an empty background. Despite the importance of this result in the context of general relativity, one must consider the more realistic case where the charged mass is imbedded into a cosmological background. Therefore, is worth to study the properties of the spacetime of a mass charged into a non-empty universe.    

McVittie\cite{mcvittie} was the first one who provided a metric for a Schwarzschild mass immersed into a cosmological background. By using isotropic coordinates, McVittie matched the Friedmann-Lemaitre-Robertson-Walker metric with the Schwarzschild metric. From a different point of view, Einstein and Straus\cite{straus} studied the problem, by cutting a spherical region in a cosmological background. In that `hole', they immersed a Schwarzschild mass and discussed the conditions to join both metrics, however they did not provide any specific metric. Considering a spatially flat Robertson-Walker cosmological fluid with pressure, Bona and Stela built their ``Swiss cheese'' model inspired in the early McVittie metric. 

By introducing charged black holes, Kastor and Traschen\cite{kastor} found a solution which describes the dynamics of a set of extremal charged black holes immersed in a de Sitter universe. More recently by using isotropic coordinates, Gao and Zhang\cite{gao} extended the McVittie solution considering a Reissner-Nordstr\"om black hole. Posada and Batic\cite{posada} studied the electrovac universe with a cosmological constant.  

In two seminal papers, Gautreau\cite{gaut1, gaut2} developed a method for imbedding a Schwarzschild mass into a cosmological background, by working with curvature coordinates $(R,T)$ and geodesic coordinates $(R,\tau)$. The Gautreau's method is based on a splitting of the spacetime in two regions, where the region inside some radius $R_{b}$ is composed of a part of cosmological fluid plus the Schwarzschild mass. Outside $R_{b}$ the energy-momentum tensor is described by a radially moving geodesic cosmological fluid. Gautreau provided examples of his imbedding method for different cosmological scenarios. An important conclusion from his work, is that planetary orbits will spiral when a Schwarzschild mass is immersed into a general universe.

In this report, we extend the Gautreau imbedding model, by considering a Reissner-Nordstr\"om charged mass. In section II we summarize the relevant expressions from references\cite{gaut1,gaut2}, particularly the metric forms and Einstein equations. The main difference will be in the energy-momentum tensor outside the region $R_{b}$ where we now have an additional contribution due to the electromagnetic tensor. In section III, we apply our results to immerse a charged mass into a de Sitter universe. In section IV we apply the Gautreau method to describe a Reissner-Nordstr\"om charged mass into an Einstein-de Sitter universe, which consists of a pressureless spacetime with $\Lambda=0$. Finally we immerse a Reissner-Nordstr\"om metric into a general universe.    

\section{Einstein equations for an imbedded mass}

In the following, we will summarize the relevant equations from references\cite{gaut1,gaut2}. In curvature coordinates $(R,T)$, the most general spherically symmetric line element that one can write is

\begin{equation}\label{curvature}
ds^2(R,T)=-A(R,T)f^{2}(R,T)dT^2+\frac{dR^2}{A(R,T)}+R^2d\Omega^{2}
\end{equation}

\noindent The physical meaning of the coordinate time $T$, is that it's recorded by clocks located at points with $R =$ const. On the other hand, one can describe the spacetime by using geodesic coordinates $(R,\tau)$ where the geodesic time $\tau$ is recorded by clocks moving along geodesics. The transformation between $T$ and $\tau$ is given by
\begin{equation}\label{trans1}
\tau_{,_{R}}=-\frac{m}{A}(k^2-A)^{1/2}
\end{equation}
\begin{equation}\label{trans2}
\tau_{,_{T}}=kf
\end{equation}
\noindent where $m=+1$ or $-1$ depending, respectively, whether the particle moves increasing or decreasing $R$. The parameter $k$ is associated to the energy per unit of mass of the particle moving along a geodesic. In terms of the coordinates $(R,\tau)$, the metric (\ref{curvature}) can be written as
\begin{equation}\label{geodesic}
ds^2(R,\tau)=-d\tau^2+\frac{1}{k^2}\left[dR-m(k^2-A)^{1/2}d\tau\right]^2+R^2d\Omega^2 
\end{equation}
\noindent We want to solve the Einstein equation
\begin{equation}\label{einstein}
R_{\nu}^{\mu}-\frac{1}{2}(R+2\Lambda)\delta_{\nu}^{\mu}=-8\pi T_{\nu}^{\mu}
\end{equation}
\noindent where we are using geometrized units $G=c=1$. From the metric (\ref{curvature}), the relevant Einstein's equations at first order in the metric components are
\begin{equation}\label{einstein1}
[R(1-A)]_{,_{R}}=-8\pi R^{2}T_{0}^{0}+R^2\Lambda
\end{equation}
\begin{equation}\label{einstein2}
[R(1-A)]_{,_{R}}-2AR\left(\frac{f_{,_{R}}}{f}\right)=-8\pi R^{2}T_{1}^{1}+R^2\Lambda
\end{equation}
\begin{equation}\label{einstein3}
[R(1-A)]_{,_{T}}=8\pi R^2T_{0}^{1}
\end{equation}
\noindent Note that these equations are valid in general for any energy-momentum tensor, which we will specify later. Given the set of equations (\ref{einstein1})-(\ref{einstein3}), Gautreau proposes to solve them by recognizing $T_{\nu}^{\mu}$ as a function of the curvature coordinates $(R,T)$, therefore they can be integrated. From (\ref{einstein1}) we can obtain then

\begin{equation}\label{int1}
A(R,T)=1+\left(\frac{8\pi}{R}\right)\int_{0}^{R}T_{0}^{0}R^2dR-\left(\frac{\Lambda}{3}\right)R^{2}
\end{equation}

\noindent from (\ref{einstein2}) by using (\ref{einstein1}) we can obtain for $f(R,T)$

\begin{equation}\label{int2}
\ln f^{2}(R,T)=8\pi\int_{0}^{R}\left(\frac{R}{A}\right)(T_{1}^{1}-T_{0}^{0})dR
\end{equation}

\noindent The component $T_{0}^{1}$ can be obtained easily from (\ref{einstein3}) 

\begin{equation}\label{int3}
T_{0}^{1}=-\frac{1}{R^2}\int_{0}^{R}T_{0_{,T}}^{0}R^2dR
\end{equation}

Instead of working with curvature coordinates, Gautreau\cite{gaut3, gaut4} showed that is advantageous to change to geodesic coordinates, which provide new features of the spacetime important in cosmology, which are unrecognized in the curvature coordinates formalism. Using the transformation (\ref{trans1}) and (\ref{trans2}), the Einstein equations (\ref{einstein1})-(\ref{einstein3}) takes the form

\begin{equation}\label{field1}
\frac{\partial}{\partial R}\left[R(1-A)\right]=-8\pi R^2\tau_{0}^{0}+R^2\Lambda
\end{equation}
\begin{equation}\label{field2}
\frac{\partial}{\partial\tau}\left[R(1-A)\right]=8\pi R^2\tau_{0}^{1}
\end{equation}
\begin{eqnarray}\label{field3}
&\frac{\partial}{\partial R}\left[R(1-A)\right]+m(k^2-A)^{-1/2}\frac{\partial}{\partial\tau}\left[R(1-A)\right]\nonumber\\
&+2m\left(\frac{RA}{k}\right)(k^2-A)^{-1/2}\left[\frac{\partial k}{\partial\tau}+m(k^2-A)^{1/2}\frac{\partial k}{\partial R}\right]\nonumber\\
&=-8\pi R^{2}\tau_{1}^{1}+R^2\Lambda 
\end{eqnarray}

\noindent where we use $\tau_{\nu}^{\mu}$ to denote the energy-momentum tensor in coordinates $(R,\tau)$. Following the notation of reference \cite{gaut2}, we use partial derivatives explicitly when we use geodesic coordinates $(R,\tau)$, to differentiate from curvature coordinates $(R,T)$ where we use commas to denote derivatives. Using the same procedure as in (\ref{int1})-(\ref{int3}), we obtain the following

\begin{equation}\label{geod1}
A(R,\tau)=1+\frac{8\pi}{R}\int_{0}^{R}R^2\tau_{0}^{0}-\frac{\Lambda}{3}R^2
\end{equation}
\begin{equation}\label{geod22}
\tau_{0}^{1}=-\frac{1}{R^{2}}\int_{0}^{R}\frac{\partial\tau_{0}^{0}}{\partial\tau}R^{2}dR
\end{equation}
\begin{eqnarray}\label{geod3}
&-m2A\left[\frac{\partial k}{\partial\tau}+m(k^2-A)^{1/2}\frac{\partial k}{\partial R}\right]=8\pi kR(k^2-A)^{1/2}\tau_{1}^{1}\nonumber\\
&-8\pi kR(k^2-A)^{1/2}\tau_{0}^{0}-m\left(\frac{8\pi k}{R}\right)\int_{0}^{R}\frac{\partial\tau_{0}^{0}}{\partial R}R^2dR 
\end{eqnarray}

Following the Gautreau's method, we break the spacetime in two regions delimited by a radius $R_{b}$. The inner region $R\leq R_{b}$ will be composed of two parts: a Reissner-Nordstr\"om charged mass (RN) and a perfect cosmological fluid

\begin{equation}\label{splitin}
\tau_{\nu}^{\mu}|_{R\leq R_{b}}=\tau_{\nu}^{\mu(RN)}+\tau_{\nu}^{\mu(cosm)}
\end{equation}

\noindent In the region $R>R_{b}$ we will have the contribution of a perfect cosmological fluid plus an electromagnetic field (em)

\begin{equation}\label{splitout}
\tau_{\nu}^{\mu}|_{R>R_{b}}=\tau_{\nu}^{\mu(cosm)}+\tau_{\nu}^{\mu(em)}
\end{equation}

\noindent The cosmological fluid is described by

\begin{equation}\label{fluid}
\tau_{\mu\nu}^{cosm}=(\rho + p)U_{\mu}U_{\nu}+pg_{\mu\nu}
\end{equation} 

\noindent where the particles are moving along radial geodesics. The non-zero components of (\ref{fluid}) are (Eqs. (2.9) reference \cite{gaut2})

\begin{equation}\label{fluid0}
\tau_{0}^{0}=-\rho
\end{equation}
\begin{equation}\label{fluid1}
\tau_{1}^{1}=p
\end{equation}
\begin{equation}\label{fluid2}
\tau_{0}^{1}=-m(k^2-A)^{1/2}(\rho+p)
\end{equation}
\begin{equation}\label{fluid3}
\tau_{2}^{2}=\tau_{3}^{3}=p
\end{equation}

\noindent In our description, we have an additional contribution due to the charge of the particle, which will be determined by the electromagnetic tensor

\begin{equation}\label{electro}
\tau^{\mu\nu(em)}=\frac{1}{4\pi}\left(F^{\mu\alpha}F_{\alpha}^{\phantom{a}\nu}+\frac{1}{4}g^{\mu\nu}F_{\alpha\beta}F^{\alpha\beta}\right)
\end{equation}
 
\noindent Consistent with the symmetry of the problem\cite{ray}, we consider only electrostatic fields, i.e., only $F^{01}\neq 0$, with the rest of components vanishing\footnote{The Maxwell equations $\nabla_{\mu}F^{\nu\mu}=4\pi J^{\nu}$ and $\nabla_{[\mu}F_{\nu\lambda]}=0$ also admits solutions with $F^{23}\neq 0$, which would indicate a magnetic monopole. However, we don't consider this situation in our treatment.}. From the relation $F_{\mu\nu}=A_{\nu_{,\mu}}-A_{\mu_{,\nu}}$ we have
\begin{equation}\label{electro1}
F_{01}=-F_{10}=-\phi_{,_{R}}
\end{equation}

\noindent From (\ref{geodesic}), (\ref{electro}) and (\ref{electro1}), the relevant non-zero components of $\tau_{\nu}^{\mu(em)}$ are

\begin{equation}\label{em0}
\tau_{0}^{0}=\tau_{1}^{1}=\frac{k^2}{8\pi}(\phi_{,_{R}})^2
\end{equation}

A worth point to mention, is that the energy parameter $k(R,\tau)$ appears explicitly in the electromagnetic tensor components. Following the procedure described in \cite{gaut2}, the field equations (\ref{geod1}) - (\ref{geod3}) in the region $R>R_{b}$ are

\begin{eqnarray}\label{important1}
A(R,\tau)&=1+\frac{8\pi}{R}\int_{0}^{R_{b}}\tau_{0}^{0}R^2dR-\frac{8\pi}{R}\int_{R_{b}}^{R}\rho R^2dR\nonumber\\
&+\frac{k^2}{R}\int_{R_{b}}^{R}(\phi_{,_{R}})^2R^2dR-\frac{\Lambda}{3}R^{2}
\end{eqnarray}
\begin{eqnarray}\label{important2}
\tau_{0}^{1}&=-\frac{1}{R^2}\int_{0}^{R_{b}}\left(\frac{\partial\tau_{0}^{0}}{\partial\tau}\right)R^2dR+\frac{1}{R^2}\int_{R_{b}}^{R}\left(\frac{\partial\rho}{\partial\tau}\right)R^2dR\nonumber\\
&-\frac{k^2}{8\pi R^2}\int_{R_{b}}^{R}(\phi_{,_{R}})^2R^2dR
\end{eqnarray}
\begin{eqnarray}\label{important3}
&\rho+p-\frac{m}{R^2}(k^2-A)^{-1/2}\left[\int_{0}^{R_{b}}\left(\frac{\partial\tau_{0}^{0}}{\partial\tau}\right)R^2dR-\int_{R_{b}}^{R}\left(\frac{\partial\rho}{\partial\tau}\right)R^2dR\right]\nonumber\\
&-\frac{m}{R^2}(k^2-A)^{-1/2}\left(\frac{k^2}{8\pi}\int_{R_{b}}^{R}(\phi_{,_{R}})^2R^2dR\right)=0
\end{eqnarray}

Now, we complete our model with the description of the energy-momentum tensor inside $R_{b}$. One contribution comes from $\tau_{0}^{0}=-\rho$, which is the cosmological fluid, and the other contribution is just the imbedded spherical mass M. In the region $R<R_{b}$ we have then

\begin{equation}\label{inner}
\tau_{0}^{0}=\tau_{0}^{0(cosm)}+M_{0}^{0}=-\rho+M_{0}^{0}
\end{equation}

\noindent where the spherical mass M is defined as

\begin{equation}\label{mass}
M=-4\pi\int_{0}^{R_{b}}M_{0}^{0}R^2dR
\end{equation}

\noindent Note that the integral that involves the electric potential $\phi$, can be rewritten in terms of the charge distribution. In that order, we recall that the potential due to a spherical charge is given by $\phi=-\frac{1}{4\pi}\frac{q}{R}$. Using this relation jointly with (\ref{inner}) and (\ref{mass}), Eqs (\ref{important1})-(\ref{important3}) becomes

\begin{equation}\label{imbed1}
A(R,\tau)=1-\frac{2M}{R}+k^2\frac{Q^2}{R^2}-\frac{8\pi}{R}\int_{0}^{R}\rho R^2dR-\frac{\Lambda}{3}R^{2}
\end{equation}
\begin{equation}\label{imbed2}
\tau_{0}^{1}=\frac{1}{R^2}\int_{0}^{R}\left(\frac{\partial\rho}{\partial\tau}\right)R^2dR-\frac{k^2}{8\pi}\frac{Q^2}{R^3}
\end{equation}
\begin{equation}\label{imbed3}
\rho+p+\frac{m}{R^2}(k^2-A)^{-1/2}\left[\int_{0}^{R}\left(\frac{\partial\rho}{\partial\tau}\right)R^2dR-\frac{k^2}{8\pi}\frac{Q^2}{R}\right]=0
\end{equation}

\noindent where we have defined $q\equiv 4\pi Q$. In the next sections, we will apply our results for the case of imbedding a charged mass into different cosmological scenarios.

\section{A Reissner-Nordstr\"om mass imbedded into a de Sitter universe}

A de Sitter spacetime corresponds to a cosmology with no matter, where the dynamics of the universe is driven by a cosmological constant. In this scenario we have $\rho=0$, such that from (\ref{imbed2}) and (\ref{imbed3}) we have

\begin{equation}\label{sitter1}
\tau_{0}^{1}=-\frac{k^2}{2R}\left(\frac{Q^2}{4\pi R^2}\right)
\end{equation}
\begin{equation}\label{sitter2}
p=\frac{mk^2}{2R}(k^2-A)^{-1/2}\left(\frac{Q^2}{4\pi R^2}\right)
\end{equation}
 
\noindent where (\ref{sitter1}) can be interpreted as flux of energy (due to the electric field) through a surface $R = $const. We also have a non-zero pressure, which can be associated to the electric field produced by the charge. Note that if we have $Q=0$, we recover the results in \cite{gaut2} as must be expected. From (\ref{imbed1}) we have

\begin{equation}\label{asitter}
A(R,\tau)=1-\frac{2M}{R}+k^2\frac{Q^2}{R^2}-\frac{\Lambda}{3}R^{2}
\end{equation}
     
\noindent and the metric (\ref{geodesic}) in geodesic coordinates $(R,\tau)$ gives

\begin{eqnarray}\label{geositter}
&ds^2(R,\tau)=-d\tau^2+R^2d\Omega^2+\nonumber\\
&\frac{1}{k^2}\left[dR-m\left(\frac{2M}{R}-k^2\frac{Q^2}{R^2}+\frac{\Lambda}{3}R^{2}+k^2-1\right)^{1/2}d\tau\right]^2
\end{eqnarray}
 
\noindent The transformation relations (\ref{trans1}) and (\ref{trans2}) takes the form

\begin{equation}\label{transitter1}
\tau_{,_{R}}=-m\frac{\left(\frac{2M}{R}-k^2\frac{Q^2}{R^2}+\frac{\Lambda}{3}R^{2}+k^2-1\right)^{1/2}}{1-\frac{2M}{R}+k^2\frac{Q^2}{R^2}-\frac{\Lambda}{3}R^{2}}
\end{equation}
\begin{equation}\label{transitter2}
\tau_{,_{T}}=k
\end{equation}

\noindent and the metric (\ref{curvature}) in curvature coordinates $(R,T)$ gives

\begin{eqnarray}\label{curvasitter}
&ds^2(R,T)=-\left(1-\frac{2M}{R}+k^2\frac{Q^2}{R^2}-\frac{\Lambda}{3}R^{2}\right)dT^2\nonumber\\
&+\left(1-\frac{2M}{R}+k^2\frac{Q^2}{R^2}-\frac{\Lambda}{3}R^{2}\right)^{-1}dR^2+R^2d\Omega^2
\end{eqnarray}

If we set $\Lambda=0$ we have the Reissner-Nordstr\"om metric, while if we set $Q=0$ we recover the Kottler metric\cite{dumin}. Therefore, the metric form (\ref{curvasitter}) can be recognized as the description of a Reissner-Nordstr\"om charged mass imbedded into a de Sitter universe. Note that if we consider the situation $Q=0$ we recover the results in \cite{gaut2}, as is expected. Note also, that the energy parameter $k(R,\tau)$ appears explicitly in \ref{curvasitter}, because we don't have a cosmological fluid to attach clocks. Once we introduce a cosmological fluid, we can consider $k=1$ even though we don't have a homogeneous situation (see argument in \cite{gaut2}). We discuss this situation in the next sections.

\section{Reissner-Nordstr\"om charged mass imbedded into an Einstein-de Sitter universe}

An Einstein-de Sitter universe corresponds to a scenario with $\Lambda=p=0$. Under this condition, (\ref{imbed3}) gives

\begin{equation}\label{eds1}
\rho+\frac{m}{R^2}(1-A)^{-1/2}\left[\int_{0}^{R}\left(\frac{\partial\rho}{\partial\tau}\right)R^2dR-\frac{1}{8\pi}\frac{Q^2}{R}\right]=0
\end{equation}

\noindent where we have set $k=1$ considering that we have a cosmological fluid to attach clocks. After substituting (\ref{imbed1}) in (\ref{eds1}) we obtain

\begin{eqnarray}\label{eds2}
&\left[\left(\frac{-m}{\rho R^2}\right)\left(\int_{0}^{R}(\frac{\partial\rho}{\partial\tau})R^2dR-\frac{1}{8\pi}\frac{Q^2}{R}\right)\right]^2=\frac{8\pi}{R}\int_{0}^{R}\rho R^2dR\nonumber\\
&+\frac{2M}{R}-\frac{Q^2}{R^2}
\end{eqnarray}
 
\noindent Following \cite{gaut2} we introduce the definition

\begin{equation}\label{zeta}
Z(R,\tau)\equiv 8\pi\int_{0}^{R}\rho R^2dR. 
\end{equation}

\noindent Using (\ref{zeta}) in (\ref{eds2}) we have

\begin{equation}\label{eds3}
R^{1/2}\left(\frac{\partial Z}{\partial\tau}-\frac{Q^2}{R}\right)+m\left[Z+\left(2M-\frac{Q^2}{R}\right)\right]^{1/2}\frac{\partial Z}{\partial R}=0
\end{equation}

\noindent and (\ref{imbed1}) takes the form

\begin{equation}\label{eds4}
A(R,\tau)=1-\left(\frac{2M+Z}{R}\right)+\frac{Q^2}{R^2}
\end{equation}
  
\noindent Note that if we set $Q=0$, results (\ref{eds2})-(\ref{eds4}) reduces to those in \cite{gaut2}. The metric (\ref{geodesic}) in coordinates $(R,\tau)$ takes the form

\begin{eqnarray}\label{geoeds}
ds^2(R,\tau)&=-d\tau^2+\left[dR-m\left(\frac{2M+Z}{R}-\frac{Q^2}{R^2}\right)^{1/2}d\tau\right]^2\nonumber\\
&+R^2d\Omega^2
\end{eqnarray}

The transformation (\ref{trans1}) between the time coordinate $T$ and the geodesic time $\tau$ becomes

\begin{equation}\label{transeds}
\tau_{,_{R}}=-m\frac{\left[\left(\frac{2M+Z}{R}\right)-\frac{Q^2}{R^2}\right]^{1/2}}{1-\left(\frac{2M+Z}{R}\right)+\frac{Q^2}{R^2}}
\end{equation}

\noindent when $R\to 0$ (\ref{transeds}) must satisfy $\tau_{,_{T}}\to 1$ (flatness condition). Finally, the metric (\ref{curvature}) in curvature coordinates takes the form

\begin{eqnarray}\label{curvaeds}
&ds^2(R,T)=-\left[1-\left(\frac{2M+Z}{R}\right)+\frac{Q^2}{R^2}\right]f^2dT^2\nonumber\\
&+\left[1-\left(\frac{2M+Z}{R}\right)+\frac{Q^2}{R^2}\right]^{-1}dR^2+R^2d\Omega^2
\end{eqnarray}

\noindent where $f(R,\tau)$ will be determined from the solution to (\ref{transeds}). In the Einstein-de Sitter universe the density is given by

\begin{equation}\label{rhoeds}
\rho=\frac{1}{6\pi\tau^2}
\end{equation}

\noindent Note that if we substitute (\ref{rhoeds}) in (\ref{zeta}) and we integrate it, after substitution in (\ref{curvaeds}) we obtain the Einstein- de Sitter universe in curvature coordinates \cite{gaut1}. That particular scenario corresponds then to a Reissner-Nordstr\"om charged mass imbedded into an Einstein-de Sitter universe. In the next section we consider a general cosmology.    

\section{Imbedding a Reissner-Nordstr\"om charged mass into a general cosmology}

We can extend our methods above, to consider the case of a charged mass imbedded into a general universe with $\rho\neq 0$ and $p\neq 0$. If we substitute (\ref{imbed1}) in (\ref{imbed3}) we obtain

\begin{eqnarray}\label{general1}
&\left[\frac{-m}{R^2}\frac{1}{(\rho+p)}\left(\int_{0}^{R}(\frac{\partial\rho}{\partial\tau})R^2dR-\frac{1}{8\pi}\frac{Q^2}{R}\right)\right]^2=\frac{8\pi}{R}\int_{0}^{R}\rho R^2dR\nonumber\\
&+\frac{2M}{R}-\frac{Q^2}{R^2}+\frac{\Lambda}{3}R^2
\end{eqnarray}

\noindent which provides a relation between density and pressure. Once we know the state equation $p=p(\rho)$, we can solve (\ref{general1}) to find $\rho(R,\tau)$. Using (\ref{zeta}) and (\ref{imbed1}), the geodesic metric (\ref{geodesic}) takes the form

\begin{eqnarray}\label{gengeo}
ds^2(R,\tau)&=\left[dR-m\left(\frac{2M+Z}{R}-\frac{Q^2}{R^2}+\frac{\Lambda}{3}R^2\right)^{1/2}d\tau\right]^2\nonumber\\
&-d\tau^2+R^2d\Omega^2
\end{eqnarray}

From (\ref{imbed2}) the flux of energy through a surface $R = $ const., gives

\begin{equation}
\tau_{0}^{1}=\frac{1}{R^2}\int_{0}^{R}\left(\frac{\partial\rho}{\partial\tau}\right)R^2dR-\frac{Q^2}{8\pi R^3}
\end{equation}

\noindent which corresponds to the energy due to the electric field (Eq. \ref{sitter1}), plus a contribution of the cosmological fluid. The transformation (\ref{trans1}) between curvature coordinates $(R,T)$ and geodesic coordinates $(R,\tau)$ takes the form

\begin{equation}\label{transgen}
\tau_{,_{R}}=-m\frac{\left[\left(\frac{2M+Z}{R}\right)-\frac{Q^2}{R^2}+\frac{\Lambda}{3}R^2\right]^{1/2}}{1-\left(\frac{2M+Z}{R}\right)+\frac{Q^2}{R^2}-\frac{\Lambda}{3}R^2}
\end{equation}

\noindent which must satisfy $\tau_{,_{T}}\to 1$ for small R. Finally, the metric for a Reissner-Nordstr\"om mass imbedded into a general universe in curvature coordinates $(R,T)$ is 

\begin{eqnarray}\label{curvagen}
&ds^2(R,T)=-\left[1-\left(\frac{2M+Z}{R}\right)+\frac{Q^2}{R^2}-\frac{\Lambda}{3}R^2\right]f^2dT^2\nonumber\\
&+\left[1-\left(\frac{2M+Z}{R}\right)+\frac{Q^2}{R^2}-\frac{\Lambda}{3}R^2\right]^{-1}dR^2+R^2d\Omega^2
\end{eqnarray}

\noindent where $f(R,\tau)$ is determined from the solution to (\ref{transgen}). Note that if we set $\Lambda=0$, our result coincides with the metric found by Gao and Zhang (Eq. 32 reference\cite{gao}). If we set $\Lambda=\rho=p=0$ we recover the Reissner-Nordstr\"om solution, while if we set $M=Q=0$ we have a cosmological metric for a general function $p=p(\rho)$. If we set $Q=0$ in (\ref{curvagen}) we recover the results of \cite{gaut2} as must be expected.  

\section{Equations of motion in geodesic coordinates}

Now that we have extended the Gautreau's imbedding method considering a Reissner-Nordstr\"om charged mass, our last step will be to provide the geodesic equations. The difference of our approach with the one developed by Gautreau\cite{gaut2}, is that we will use the geodesic metric form (\ref{geodesic}), while Gautreau used the curvature form (\ref{curvature}). The geodesic equations are given by

\begin{equation}\label{eom}
\frac{dV^{\mu}}{ds}+\Gamma_{\rho\sigma}^{\mu}V^{\rho}V^{\sigma}=0
\end{equation}
\noindent where $V^{\mu}=\frac{dx^{\mu}}{ds}$, and we take $s$ as the proper time along a particle's trajectory. From (\ref{geodesic}), the relevant affine connections are\footnote{We used the \emph{ctensor} package available in \emph{Maxima} to calculate the Christoffel symbols. Special thanks to Viktor T. Toth.}

\begin{equation}
\Gamma_{00}^{0}=-\frac{m}{2}(1-A)^{1/2}\frac{\partial A}{\partial R}\,\,\,\, ; \,\,\,\, \Gamma_{10}^{1}=-\Gamma_{00}^{0}\nonumber  
\end{equation}
\begin{equation}
\Gamma_{00}^{1}=-\frac{1}{2}\left[A\frac{\partial A}{\partial R}-\frac{m}{(1-A)^{1/2}}\frac{\partial A}{\partial\tau}\right]\,\,\,\, ; \,\,\,\,\Gamma_{10}^{0}=\frac{1}{2}\frac{\partial A}{\partial R}\nonumber
\end{equation}
\begin{equation}\label{christo}
\Gamma_{11}^{0}=-\frac{m}{2}(1-A)^{-1/2}\frac{\partial A}{\partial R}\,\,\,\,\, ; \,\,\,\,\, \Gamma_{11}^{1}=-\Gamma_{10}^{0}
\end{equation}
\begin{equation}
\Gamma_{12}^{2}=\Gamma_{13}^{3}=\frac{1}{R}\,\,\,\, ; \,\,\,\, \Gamma_{23}^{3}=\cot\theta\,\,\,\, ; \,\,\,\, \Gamma_{33}^{2}=-\sin\theta\cos\theta\nonumber
\end{equation}
\begin{equation}
\Gamma_{22}^{0}=mR(1-A)^{1/2}\,\,\,\,\, ; \,\,\,\,\, \Gamma_{33}^{0}=\Gamma_{22}^{0}\sin^2\theta\nonumber
\end{equation}
\begin{equation}
\Gamma_{22}^{1}=-RA\nonumber\,\,\,\,\, ; \,\,\,\,\, \Gamma_{33}^{1}=\Gamma_{22}^{1}\sin^2\theta\nonumber
\end{equation} 

\noindent We can specialize to the case $\theta=\frac{\pi}{2}$ such that $V^2=0$. Using (\ref{christo}) in (\ref{eom}) we have

\begin{eqnarray}\label{eom1}
&\frac{dV^0}{ds}-\frac{m}{2}(1-A)^{1/2}\frac{\partial A}{\partial R}\left[(V^{0})^2+(V^{1})^2\right]+\frac{\partial A}{\partial R}(V^{0}V^{1})\nonumber\\
&+mR(1-A)^{1/2}(V^{3})^2=0
\end{eqnarray}
\begin{eqnarray}\label{eom2}
&\frac{dV^1}{ds}+\frac{1}{2}\left[A\frac{\partial A}{\partial R}-\frac{m}{(1-A)^{1/2}}\frac{\partial A}{\partial\tau}\right](V^0)^2-\frac{1}{2}\frac{\partial A}{\partial R}(V^1)^2\nonumber\\
&+\frac{m}{2}(1-A)^{1/2}\frac{\partial A}{\partial R}(V^0V^1)-RA(V^3)^2=0
\end{eqnarray}
\begin{eqnarray}\label{eom3}
\frac{dV^3}{ds}+\frac{2}{R}(V^1V^3)=0
\end{eqnarray}
\begin{eqnarray}\label{eom4}
&-A(V^0)^2-2m(1-A)^{1/2}V^0V^1+(V^1)^2\nonumber\\
&+R^2(V^3)^2=-1
\end{eqnarray}

\noindent where (\ref{eom4}) is just the normalization condition $g_{\mu\nu}V^{\mu}V^{\nu}=-1$. We notice that (\ref{eom3}) can be rewritten as

\begin{equation}\label{angular}
R^2\frac{dV^3}{ds}+2RV^3\frac{dR}{ds}=\frac{d}{ds}(R^2V^3)=0
\end{equation} 

\noindent which implies

\begin{equation}\label{angmom}
R^2V^3=R^2\frac{d\phi}{ds}=L=const.
\end{equation} 

\noindent where $L$ is the angular momentum per unit of mass. Note that (\ref{angmom}) is just the angular momentum conservation. One important result found by Gautreau by using curvature coordinates\cite{gaut2}, is that planetary orbits will spiral when a Schwarzschild mass is imbedded into a cosmological background. In the following we arrive to the same conclusion, with the difference that we work in geodesic coordinates $(R,\tau)$. Let's assume a planetary circular motion, i.e.,

\begin{equation}\label{circ}
R=const. \Rightarrow \frac{dR}{ds}=0
\end{equation} 
  
\noindent We will show that this condition can't be satisfied if we want to satisfy the EOM (\ref{eom1})-(\ref{eom4}). Therefore $R\neq $const., which implies that there is a spiralling of orbits. Using condition (\ref{circ}) we can write

\begin{equation}\label{total}
\frac{dV^0}{ds}=\frac{dV^0}{d\tau}\frac{d\tau}{ds}=V^0\frac{dV^0}{d\tau}
\end{equation} 

\noindent using (\ref{total}) in (\ref{eom1}) we have

\begin{eqnarray}\label{circ1}
&V^0R^2\frac{dV^0}{d\tau}-\frac{m}{2}R^2(1-A)^{1/2}\frac{\partial A}{\partial R}(V^{0})^2\nonumber\\
&+mR(1-A)^{1/2}\left[A(V^0)^2-1\right]=0
\end{eqnarray}

\noindent We can rewrite (\ref{eom4}) in terms of $L$ like

\begin{equation}\label{circ2}
A(V^0)^2=1+\left(\frac{L}{R}\right)^2
\end{equation} 

\noindent Using (\ref{circ2}) in (\ref{eom2}) we obtain

\begin{eqnarray}\label{orbit}
\frac{\partial}{\partial R}\left[\ln A(R,\tau)\right]-\frac{m}{A}(1-A)^{-1/2}\frac{\partial}{\partial\tau}\left[\ln A(R,\tau)\right]\nonumber\\
=\frac{2/R}{\left[1+(R/L)^2\right]}
\end{eqnarray} 

We started with the condition $R = $ const., which makes the right side of (\ref{orbit}) constant, however if $A$ is function of $\tau$, the left side is not going to be constant. This contradiction leads us to conclude that the condition $R = $ constant, is not consistent. Only for the case of a Reissner-Nordstr\"om-de Sitter spacetime (RNS)(\ref{asitter}), where $A$ is not a function of $\tau$, the orbit radius will be constant. However, in the most general case where $A=A(R,\tau)$ the radius of the orbit will change. Note that for the (RNS) case, the second term to the left of (\ref{orbit}) vanishes, which reduces to the expression found by Gautreau (Eq. (6.12), reference \cite{gaut2}). Therefore, we can consider (\ref{orbit}) as the generalization of the equation for the orbit spiralling, in geodesic coordinates. Once we choose a cosmological model, we can determine $A(R,\tau)$ and use (\ref{orbit}) to find the equation for the radius of the orbit.    

In order to provide a magnitude of the spiraling of an orbit, Gautreau makes an estimate using Newtonian physics (Eq. (6.26b) reference \cite{gaut2}). From his calculation, Gautreau concludes that changes in the orbital radius at the scale of our solar system are negligible, but these becomes appreciable at the scale of $R\approx 25 Kpc$. Criticism can be raised about his development. In principle, his Newtonian approximation is unrelated to his imbedding formalism, which provides a new way to consider cosmological theory. A more realistic and consistent calculation, would involve to solve the EOM (\ref{eom1})-(\ref{eom4}) in order to obtain an equation for $\frac{dR}{d\tau}$. We see the high complexity of the equations due to the mixed terms $(V^0V^1)$, which makes a hard task to solve them. Instead of that, we can use the curvature coordinates $(R,T)$ which provides more simple EOM (see Eqs. (6.3)-(6.5) reference \cite{gaut2}). However, the metric component $B(R,T)$ depends on the function $\tau_{,_{T}}$, which is not easy to solve in principle. If I am allowed to speculate here, I suggest that using numerical methods might help to find approximate solutions. This problem will be left for future investigation.   

\section{Discussion}

We have extended the Gautreau's imbedding method, by considering a Reissner-Nordstr\"om charged mass. We worked in curvature coordinates $(R,T)$ where $T$ is associated to times recorded by clocks at fixed points $R = $ const. However, we can also work cosmological theory by introducing geodesic coordinates $(R,\tau)$, where $\tau$ is the time recorded by clocks radially moving along geodesics with the cosmological fluid. One important consequence of our description, is that when we immerse a mass into a cosmological background, planetary orbits will spiral. We found that only for the (RNS) field, where the metric component $A$ is independent of $\tau$, the orbit radius will be constant. However, in a more general cosmological scenario, the orbit radius will change.

We raised some criticism to the equation for the orbit change found by Gautreau, considering that his approach involved a Newtonian calculation, which is not consistent with the imbedding formalism developed previously. In order to obtain an equation for the orbital change, which will be more consistent with the formalism developed, it's necessary to solve the EOM (\ref{eom1})-(\ref{eom4}). The complexity of these equations, will require perhaps, to use numerical methods in order to find approximate solutions. That problem is left open for future investigation.  

\section{Acknowledgments}

I want to thank to the Department of Physics and Astronomy at the University of South Carolina for its support while this work was developed.

\end{document}